\documentclass[12pt,a4paper]{article}
\usepackage{amsmath,graphics,epsfig,rotating,cite}

\begin{document}

\setlength{\unitlength}{1mm}

\def\mbf             {\boldmath}
\def\ti              {\tilde}

\def\a               {\alpha}
\def\e               {\varepsilon}
\def\b               {\beta}
\def\d               {\delta}
\def\D               {\Delta}
\def\g               {\gamma}
\def\G               {\Gamma}
\def\l               {\lambda}
\def\t               {\theta}
\def\s               {\sigma}
\def\S               {\Sigma}
\def\x               {\chi}

\def\ch              {\ti \x^\pm}
\def\nt              {\ti \x^0}
\def\sg              {\ti g}
\def\st              {\ti t}
\def\sb              {\ti b}

\newcommand{\mst}[1]   {m_{\ti t_{#1}}}
\newcommand{\msb}[1]   {m_{\ti b_{#1}}}
\newcommand{\mstau}[1] {m_{\ti\tau_{#1}}}
\newcommand{\mch}[1]   {m_{\ti \x^\pm_{#1}}}
\newcommand{\mnt}[1]   {m_{\ti \x^0_{#1}}}
\newcommand{\msg}      {m_{\ti g}}

\def\MS                {{\rm \overline{MS}}}
\def\DR                {{\rm \overline{DR}}}

\newcommand{\isajet}     {{\tt Isajet\,7.72}}
\newcommand{\isajetNew}   {{\tt Isajet\,7.73}}
\newcommand{\softsusy}   {{\tt SoftSusy\,2.0}}
\newcommand{\spheno}     {{\tt SPheno\,2.2.3}}
\newcommand{\suspect}    {{\tt Suspect\,2.3.4}}

\newcommand{\isajetnn}     {{\tt Isajet}}
\newcommand{\softsusynn}   {{\tt SoftSusy}}
\newcommand{\sphenonn}     {{\tt SPheno}}
\newcommand{\suspectnn}    {{\tt Suspect}}

\newcommand{\eq}[1]  {\mbox{(\ref{eq:#1})}}
\newcommand{\fig}[1] {Fig.~\ref{fig:#1}}
\newcommand{\Fig}[1] {Figure~\ref{fig:#1}}
\newcommand{\tab}[1] {Table~\ref{tab:#1}}
\newcommand{\Tab}[1] {Table~\ref{tab:#1}}
\newcommand{\sect}[1] {Sect.~\ref{sect:#1}}
\newcommand{\Sect}[1] {Section~\ref{sect:#1}}

\newcommand{\change}     {\marginpar{\bf change}}

\newcommand{\gsim}{\;\raisebox{-0.9ex}
           {$\textstyle\stackrel{\textstyle >}{\sim}$}\;}

\newcommand{\lsim}{\;\raisebox{-0.9ex}{$\textstyle\stackrel{\textstyle<}
           {\sim}$}\;}

\newcommand{\smaf}[2] {{\textstyle \frac{#1}{#2} }}


\vspace*{-30mm}
\begin{flushright}
  CERN-PH-TH/2005-217\\ FSU-HEP-051107 \\
  IFIC/05-57\\ LBNL-59093\\ ZU-TH 20/05\\
  hep-ph/0511123
\end{flushright}
\vspace*{4mm}

\begin{center}
{\Large\bf  On the treatment of threshold effects\\[3mm] 
            in SUSY spectrum computations} \\[8mm]

{\large  H. Baer$^{\,1}$, J. Ferrandis$^{\,2}$, 
S. Kraml$^{\,3}$, W. Porod$^{\,4,5}$ }\\[4mm]

{\it 1) Dept.\ of Physics, Florida State University, 
        Tallahassee, FL 32306, USA\\[1mm]
     2) Lawrence Berkeley Laboratory, Berkeley, CA 94720, USA\\[1mm]
     3) Theory Division, Dept.\ of Physics, CERN, CH-1211 Geneva 23, Switzerland\\[1mm]
     4) Instituto de F\'isica Corpuscular, CSIC, Val\`encia, Spain\\[1mm]
     5) Institut f\"ur Theoretische Physik, Univ. Z\"urich, Switzerland }\\[8mm]

\end{center}

\begin{abstract}
We take a critical view of the treatment of threshold effects 
in SUSY spectrum computations from high-scale input. 
We discuss the two principal methods of (a)~renormalization at a common 
SUSY scale versus (b)~integrating out sparticles at their own mass scales.
We point out problems in the implementations in public spectrum codes,
together with suggestions for improvements.   
In concrete examples, we compare results of \isajet\ and \spheno, and 
present the improvements done in \isajetNew. 
We also comment on theoretical uncertainties. 
Last but not least, we outline how a consistent 
multiscale approach may be achieved.
\end{abstract}

\section{Introduction}

It is often argued \cite{Blair:2000gy,Allanach:2004ud,Weiglein:2004hn,spa} 
that, from measurements of supersymmetry (SUSY) at the LHC and ILC, 
the model parameters can be extracted precisely enough   
to test the high-scale structure of the theory, 
in other words to test the boundary conditions of the soft SUSY breaking
(SSB) terms.  
This requires relating, with very high precision, sparticle properties 
measured at the TeV energy scale with the Lagrange parameters at the 
high-energy scale, often the GUT scale.  

Of course, the question arises as to what theoretical uncertainties are 
involved in this exercise. Such uncertainties originate 
from truncating the perturbation series of 
(a)~the running of the $\DR$ parameters between the 
electroweak (EW) and the high-energy scale, and 
(b)~the relation at the EW scale between $\DR$ and on-shell SUSY parameters. 
They have been investigated in \cite{Allanach:2003jw} 
by comparing different state-of-the-art spectrum computations. 
Differences at the level of a few per cent have been found, 
part of which have been traced to higher-order loop effects.  
Since then important improvements
have been made in all codes. 
When comparing today the latest versions of 
\isajetnn\ \cite{Paige:2003mg},
\softsusynn\ \cite{Allanach:2001kg},
\sphenonn\ \cite{Porod:2003um} and 
\suspectnn\ \cite{Djouadi:2002ze}, 
the typical spread in the results is $\lsim 1\%$,  
which is compatible with the expected precisions at the LHC and getting 
close to those expected at the planned International Linear Collider (ILC)  
(see \cite{online} for an online comparison). 

One exception is the mass of the lightest neutralino.
On the one hand, $\mnt{1}$ is expected to be measured with per-mille 
precision at the ILC. On the other hand, in mSUGRA with a bino-like LSP, 
one finds differences of few per cent in the $\mnt{1}$ obtained from 
\isajet\ as compared to $\mnt{1}$ obtained from  \softsusy, \spheno\ 
or \suspect\ (these latter three programs typically agree to $\sim 0.5\%$ 
on $\mnt{1}$). 
In this context it should also be noted that 
a $\sim1$~GeV error in the mass of the LSP can translate into a 
$\sim 10\%$ error in the prediction of its relic density 
\cite{Allanach:2004xn}.

The public spectrum codes under consideration all have 2-loop 
renormalization group (RG) running implemented. 
A major difference
between \isajetnn\ on the one side and 
\softsusynn, \sphenonn\ and \suspectnn\ on the other lies
in the treatment of threshold effects 
for computing the sparticle pole masses. 
This is the topic of this paper. 
It was the above mentioned discrepancy in the prediction of the 
LSP mass that motivated this study. 
Our analysis, however, applies to the computation of the pole masses
of mixing sparticles in general. 

The paper is organized as follows. In Sect.~2 we briefly explain the 
two methods used to compute the sparticle pole masses. In Sect.~3 we 
review these methods for the example of the neutralino sector. We compare 
results of \spheno\ and \isajet, and discuss some shortcomings 
in the computations as well as ways of improving. In Sect.~4 we then 
analyse the scalar top sector.  
An improved scheme for the computation of sparticle masses in \isajetnn\ 
is presented in Sect.~5.  
The technical difficulties involved with a consistent multiscale approach
are discussed in more detail in Sect.~6, and finally 
Sect.~7 contains our conclusions.

\section{Two methods}

The RG equations are employed in the $\DR$ scheme. The SUSY mass 
parameters obtained from the RG evolution are hence $\DR$ running ones. 
In order to obtain the physical sparticle masses one therefore has 
to perform the proper shifts to the on-shell scheme.
The public 
SUSY spectrum codes can in fact be classified by 
their method of determining the sparticle pole masses. 

One approach, adopted by \softsusynn, \sphenonn\ and \suspectnn\footnote{Early 
versions of \suspectnn\ used the step-beta function approach described below; 
the method of a common renormalization scale is used from version 2.0 
onwards.}, is to assume that the full set of MSSM RGEs (gauge and Yukawa 
couplings plus soft terms) are valid between the scales $M_Z$ and $M_{\rm GUT}$.
The gauge and Yukawa coupling boundary conditions are stipulated at
$Q=M_Z$, while soft term boundary conditions are stipulated
at $Q=M_{\rm GUT}$. The RGEs are run iteratively between $M_Z$ and $M_{\rm GUT}$
until a convergent solution is found. The $\DR$ parameters are extracted 
at a common renormalization scale $Q$, 
usually taken to be 
$Q=M_{\rm SUSY}=\sqrt{\mst{L}\mst{R}}$ or $\sqrt{\mst{1}\mst{2}}$.
The one-loop (logarithmic and finite) self-energy corrections 
\cite{Pierce:1996zz} are then added at that scale \footnote{For the neutral
Higgs masses and the $\mu$ parameter, also the two-loop corrections of
 \cite{Degrassi:2001yf} are included.}.
We will refer to this method as `common scale approach'.
It is relatively straightforward and certainly self-consistent, but misses 
two-loop logarithmic contributions between the renormalization scale 
$Q$ and the actual mass scale of the sparticle. 
Such logarithmic corrections could be relevant in cases where the mass scales
involved are severely split, such as in focus-point supersymmetry.

The other method, proposed in \cite{Castano:1993ri,Dedes:1995sb}, 
is to adopt a multiscale effective theory approach, and to functionally 
integrate out all heavy degrees of freedom at each particle threshold. 
As this is realized by the use of a theta function at each threshold, 
we call it the `step-beta function approach' 
(implying continuous matching conditions for the remaining parameters).
The program \isajetnn\ does adopt a hybrid technique along these lines.
In \isajet, the full two-loop MSSM RGEs are employed between
$M_Z$ and $M_{\rm GUT}$, with the exception that one-loop step-beta functions  
are adopted for gauge and Yukawa couplings. (This approach means that log
corrections to gauge and Yukawa couplings at the  scale $Q=M_Z$ are not 
needed, since they are handled by the RG evolution). 
In addition, each SSB parameter $m_i$ is extracted from the RGEs at an 
energy scale equal to 
$m_i=m_i(m_i)$, with the exception of 
parameters involved in the Higgs potential, which are all extracted at the
common scale $Q=M_{\rm SUSY}=\sqrt{\mst{L}\mst{R}}$.

The RG evolution of mass parameters from high to low scales 
is equivalent to computing the logarithmic radiative corrections.
The step-beta function approach hence gives a leading-log approximation 
of $m_i=m_i(m_i)$. 
For sparticles that do not mix, i.e.\ gluinos as well as squarks and 
sleptons of the first and second generations, this 
approach directly gives 
the pole masses up to constant terms. For the required level of precision 
these constant terms are important, so they have to be added at the end 
of the running.  
The situation is more complicated for mixing sparticles, since multiple
mass scales can be involved in the mixing matrices, 
and care has to be taken to compute the on-shell mass matrices  
in a consistent way. This will be discussed in more detail in the 
following sections, first for the example of the neutralino sector, and then 
for the case of scalar tops.  
Moreover, as we will discuss in \sect{discussion}, at the two-loop level,  
each time a sparticle is integrated out this implies non-trivial matching 
conditions for the parameters remaining in the effective theory. 
To our knowledge these matching conditions are not yet known and, thus,
are not yet taken into account properly. There are also other 
complications, which we will discuss in \sect{discussion}.

\section{Neutralino sector}

\subsection{Neutralino mass matrix}

At lowest order, the neutralino mass matrix in the basis 
$\psi_j^0=(-i\lambda ',-i\lambda^3,\psi_{H_1}^0,\psi_{H_2}^0)$ 
is  
\begin{equation}
  {\cal M}_N =
  \left( \begin{array}{cccc}
  M_1 & 0 & -m_Z s_W c_\beta  & m_Z s_W s_\beta \\
  0 & M_2 &  m_Z c_W c_\beta  & -m_Z c_W s_\beta  \\
  -m_Z s_W c_\beta & m_Z c_W c_\beta   & 0 & -\mu \\
   m_Z s_W s_\beta & - m_Z c_W s_\beta & -\mu & 0
  \end{array}\right)
  \label{eq:ntmassmat}
\end{equation}
with $s_W=\sin\theta_W$, $c_W=\cos\theta_W$, $s_\beta=\sin\beta$, 
$c_\beta=\cos\beta$ and $\tan\beta = v_2/v_1$. 
It is diagonalized by a unitary mixing matrix $N$: 
\begin{equation}
  N{{\cal M}_N} N^T =
  {\rm diag}(\e_1\mnt{1},\,\e_2\mnt{2},\,
             \e_3\mnt{3},\,\e_4\mnt{4})\,,
\end{equation}
where $\mnt{i}$ $(i=1,...,4)$ are the neutralino masses,   
$\mnt{1}<....<\mnt{4}$, and $\e_i$ are their signs. The mass eigenstates 
are $\nt_i=N_{ij}\psi_j^0$. 

The SUSY parameters taken out of the RG running are $\DR$ parameters 
at a certain scale $Q$. Straightforward diagonalization of Eq.~\eq{ntmassmat} 
therefore gives the $\DR$ running mass eigenvalues. 
In order to obtain the neutralino pole masses, one has to add 
self-energy corrections  
\begin{equation}
  {\cal M}_N^{\rm onshell} = {\cal M}_N(Q) + \D {\cal M}_N(Q)\,, 
  \label{eq:mntcorr}  
\end{equation}
leading to corrections in the masses, $\mnt{i}\to\mnt{i}+\D\mnt{i}$,
and in the mixing matrix $N \to N+\D N$. 
Here notice that the on-shell condition $p^2=m^2$ has to be fulfilled 
for each $\mnt{i}$ separately; so one has to diagonalize 
Eq.~(\ref{eq:mntcorr}) four times\footnote{This actually leads to a 
(numerically very small) ambiguity in the neutralino mixing matrix. 
In \isajetnn\,{\tt 7.69-7.72}, tree-level mixing elements are adopted 
for cross section calculations, along with loop-corrected neutralino masses. 
In \spheno, the mixing matrix obtained for $p^2=\mnt{1}^2$ is used.}.
The corrections at the one-loop level are given in \cite{Pierce:1996zz}; 
see also the discussion in \cite{Oller:2003ge}. 
They typically amount to a few per cent.
Since the shift from the $\DR$ to the on-shell sheme, Eq.~\eq{mntcorr}, 
cannot be performed 
to all orders, there will always be a small residual scale dependence 
of the pole masses. 
This scale dependence is often regarded as an estimate of higher-order 
corrections.

In the step-beta function approach, the mass parameters that enter 
Eq.~\eq{ntmassmat} are $M_1(M_1)$, $M_2(M_2)$, and $\mu(M_{\rm SUSY})$.  
According to \cite{Dedes:1995sb}, this  
corresponds to the on-shell ${\cal M}_N$ up to finite corrections. 
The on-shell mass matrix at the full one-loop level is then given by 
\begin{equation}
  {\cal M}_N^{\rm onshell} = {\cal M}_N^{\rm log.corr} + 
                             \D {\cal M}_N^{\rm const}\,. 
  \label{eq:mntcorr2}  
\end{equation}
The complications involved in this procedure will be discussed in 
\sect{discussion}.\\

\subsection{Threshold corrections at a common scale}

\begin{figure}[t]
  \centerline{\epsfig{file=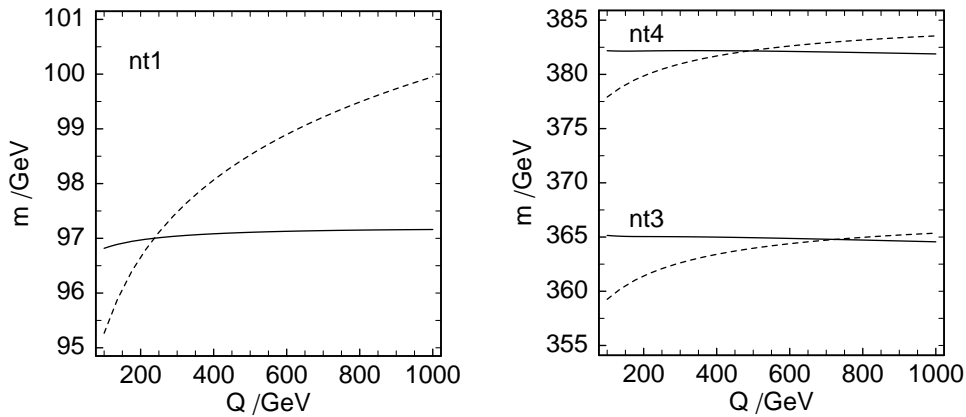,height=6.4cm}}
  \caption{Dependence of the masses of $\nt_1$ (left) and $\nt_{3,4}$ 
           (right) on the renormalization scale $Q$. The dashed lines 
           are the $\DR$ running masses, the full lines are the one-loop 
           corrected pole masses. Computed with \spheno.}
\label{fig:mnt-spheno}
\end{figure}

We first discuss results 
for the approach of running all SUSY-breaking parameters to a common 
renormalization scale $Q$ and adding one-loop threshold corrections 
at this scale. As already mentioned, this is the method employed in 
\softsusynn, \sphenonn\ and \suspectnn. 
We use the mSUGRA benchmark point SPS1a with 
\begin{equation}
   m_0=100,\quad m_{1/2}=250,\quad A_0=-100,\quad \tan\b=10, 
   \quad \mu>0, \quad m_t=175 
\label{eq:sps1a}
\end{equation}
as an illustrative numerical example. 
Moreover, we take $\alpha_s(M_Z)=0.1172$, and $m_b(m_b)=4.214$~GeV 
as SM input values in the $\MS$ scheme so as to be able to compare 
with \isajet.

The dependence of the $\DR$ running (tree-level) neutralino masses 
and the one-loop corrected pole masses on the scale $Q$ is shown 
in \fig{mnt-spheno}, for $Q=100$~GeV to 1~TeV. 
In addition, \tab{mnt-spheno} lists the $\DR$ mass parameters 
together with the tree-level and one-loop corrected  
neutralino masses at SPS1a for some particular 
choices of $Q$. The numbers have been obtained with \spheno, 
which has the complete one-loop corrections of \cite{Pierce:1996zz} 
for all sparticle masses. 
As can be seen, going from the $\DR$ to the on-shell scheme is 
a quite important correction. 
The scale dependence of the 
$\DR$ (or tree-level) neutralino masses, $\mnt{i}^{\DR}$, is about 5~GeV 
between $Q=100$~GeV and $Q=1$~TeV. That is about 5\% for the LSP 
mass and about 2--3\% for the masses of the heavier neutralinos. 
For the one-loop corrected pole masses, $\mnt{i}^{\rm pole}$, 
the scale dependence goes down to the level of a few per mille. 
Here notice also that the scale dependence 
is largest for the wino-like $\nt_2$. 

The results of \softsusy\ and \suspect\ are very similar to those of \spheno. 
For illustration the results obtained with \suspect\ are listed 
in \tab{mnt-suspect}. Here
the scale dependence of  $\mnt{2,3}^{\rm pole}$ is
slightly larger, because the self-energy corrections are applied 
in the approximation given in Sect.~4.2 in \cite{Pierce:1996zz}.

The scale dependence gives an estimate of the size of the missing 
higher-order logarithmic corrections. 
This is not necessarily the full theoretical uncertainty.
One point of caution is, for instance,  
the fact the SUSY threshold corrections to the gauge and Yukawa couplings 
are computed collectively at $M_Z$. 
This is a valid choice. However, the matching between SM and MSSM couplings 
could also be done at, for instance, $m_{\rm LSP}$ or $M_{\rm SUSY}$. 
The thus induced uncertainty is not taken into account by the
scale dependence as it results in a shift in the boundary conditions.

\begin{table}[t]
\begin{center}
\begin{tabular}{c||ccc|cccc|cccc}
  $Q$ & $M_1$ & $M_2$ & $\mu$
   & $\mnt{1}^{\DR}$ & $\mnt{2}^{\DR}$ & $\mnt{3}^{\DR}$ & $\mnt{4}^{\DR}$
   & $\mnt{1}^{\rm pole}$ & $\mnt{2}^{\rm pole}$ 
   & $\mnt{3}^{\rm pole}$ & $\mnt{4}^{\rm pole}$\\
\hline
  100        & 98.39 &  188.9 &  352.7 & 95.27 &  173.3 &  359.3 &  377.9  & 96.82 &  179.9 &  365.2 &  382.2 \\
  200        & 99.78 &  190.1 &  355.0 & 96.66 &  174.8 &  361.4 &  379.9 & 96.97 &  180.3 &  365.0 &  382.2 \\
  $M_{\rm SUSY}$ & 101.6 &  191.7 &  357.6 & 98.47 &  176.6 &  363.9 &  382.2 & 97.11 &  180.7 &  364.9 &  382.2 \\
  1000       & 103.1 &  193.0 &  359.2 & 99.95 &  178.0 &  365.4 &  383.5 & 97.16 &  181.0 &  364.6 &  381.9 \\
\hline
\end{tabular}\end{center}
\caption{Parameters and masses (in GeV) for the mSUGRA benchmark point SPS1a
  obtained with \spheno\ for different renormalization scales $Q$;
  $M_{\rm SUSY}=484.5$~GeV.}
\label{tab:mnt-spheno}
\end{table}

\begin{table}[t]
\begin{center}
\begin{tabular}{c||ccc|cccc|cccc}
  $Q$ & $M_1$ & $M_2$ & $\mu$ 
   & $\mnt{1}^{\DR}$ & $\mnt{2}^{\DR}$ & $\mnt{3}^{\DR}$ & $\mnt{4}^{\DR}$
   & $\mnt{1}^{\rm pole}$ & $\mnt{2}^{\rm pole}$ 
   & $\mnt{3}^{\rm pole}$ & $\mnt{4}^{\rm pole}$\\
\hline
  100        & 98.55& 188.9& 352.6 & 95.25& 173.3& 358.5& 377.2 & 97.00& 179.7& 363.4& 382.4\\
  200        & 99.96& 190.2& 354.8 & 96.66& 174.7& 360.5& 379.1 & 97.18& 180.2& 363.0& 381.9\\
  $M_{\rm SUSY}$ & 101.5& 191.6& 356.7 & 98.38& 176.4& 363.0& 381.3 & 97.31& 180.8& 362.8& 381.7\\
  1000       & 103.3& 193.1& 358.9 & 99.96& 177.9& 364.4& 382.7 & 97.37& 181.2& 362.1& 381.0\\
\hline
\end{tabular}\end{center}
\caption{Same as \tab{mnt-spheno}, but computed with \suspect;
         $M_{\rm SUSY}=465.4$~GeV.}
\label{tab:mnt-suspect}
\end{table}

\subsection{Step-beta functions in Isajet}

  In the implementation employed in \isajet , each SSB parameter 
(aside from Higgs potential parameters) is 
extracted from the RG running at the scale equal to the SSB mass value.
In particular, the parameters in the neutralino sector are 
$M_1(M_1)$, $M_2(M_2)$, $\mu(M_{\rm SUSY})$ and $\tan\b(M_{\rm SUSY})$
\footnote{Inside \isajetnn, $M_{\rm SUSY}$ is called  {\tt HIGFRZ} and given by\\ 
{\tt HIGFRZ=SQRT(MAX(AMZ**2,AMTLSS*AMTRSS*SIGN(1.,AMTLSS*AMTRSS)))}.}.
In this way logarithmic threshold corrections are included, and
diagonalizing the tree-level mass matrix of Eq.~\eq{ntmassmat} gives 
a leading-log approximation of the neutralino pole masses. 
Finite corrections, however, 
can be of the same order as 
the logarithmic ones, so they have to be taken into account according to 
Eq.~\eq{mntcorr2}.

At this point some comments are in order on the actual implementation 
of the step-beta function approach in \isajetnn. 
Up to and including \isajetnn\ version {\tt 7.72}:
\begin{itemize}
\item While the SSB parameters are extracted from the RG running 
      at their respective mass scale, the SSB parameters are not formally
      `integrated out', so that the soft term RGEs remain those of the MSSM
      all the way from $M_{\rm GUT}$ to $M_Z$ (unlike the case of the 
      gauge and Yukawa couplings, where the beta functions change 
      each time a threshold is passed). 
      Thus, the \isajetnn\ algorithm is actually a mixed scheme.
\item For the finite shifts, the full expressions of \cite{Pierce:1996zz} 
      for the one-loop self-energies are used. These involve $A_0$, $B_0$ 
      and $B_1$ functions, which depend on the renormalization scale $Q$ 
      and are evaluated at $Q=M_{\rm SUSY}$ in \isajetnn\,{\tt 7.69-7.72}. 
      This leads to a double counting of logs between $M_{\rm SUSY}$ and the 
      actual mass scale of the sparticle.
\end{itemize}
%
Numerically \isajet\ agrees quite well with the other 
public codes. The exception is the LSP mass, which,  
as already mentioned in the introduction, typically turns out to be
a few per cent smaller than the results from 
\softsusy, \suspect\ and \spheno. This can be understood as an effect of the 
above mentioned double counting, which induces an error 
$\propto {\rm ln}(\ti m^2/M_{\rm SUSY}^2)$, where $\ti m$ is the mass parameter 
of the sparticle considered. Obviously, the effect is largest for the LSP.

\begin{table}\begin{center}
\begin{tabular}{c||ccc|cccc|cccc}
  Case & $M_1$ & $M_2$ & $\mu$ 
   & $\mnt{1}^{(0)}$ & $\mnt{2}^{(0)}$ & $\mnt{3}^{(0)}$ & $\mnt{4}^{(0)}$
   & $\mnt{1}^{(1)}$ & $\mnt{2}^{(1)}$ & $\mnt{3}^{(1)}$ & $\mnt{4}^{(1)}$\\
\hline
  A &  99.5& 192.4& 351.2 & 96.2& 176.4& 358.0& 377.0 & 95.2& 180.5& 357.0& 377.5\\
  B &  99.5& 192.4& 351.2 & 96.2& 176.4& 358.0& 377.0 & 97.9& 182.0& 357.6& 377.9\\
  C & 103.2& 192.9& 345.1 & 99.9& 176.4& 351.5& 371.4 & 101.5& 181.7& 351.6 
& 372.6\\
\hline
  D & 101.7& 192.1& 350.9 & 98.4& 176.2& 357.9& 376.9 & 97.3& 180.2& 356.7& 377.2\\
\hline
\end{tabular}\end{center}
\caption{\isajetnn\ results (in GeV) for the neutralino sector at SPS1a.  
  Case A is the original \isajet;
  Case B is \isajet\ with the improvement that the one-loop self-energies  
  are each computed at their relevant scale as explained 
  in the text; and  
  Case C employs step-beta functions for all SUSY parameters.   
  In case D, the SUSY parameters are all frozen out at 
  $M_{\rm SUSY}=456$~GeV, and the one-loop corrections are applied at 
  this scale.}
\label{tab:mnt-isajet}
\end{table}

As a concrete example, the results of \isajet\ for the parameters of 
SPS1a, Eq.~\eq{sps1a}, are given as Case~A in \tab{mnt-isajet}. 
Here $\mnt{i}^{(0)}$ corresponds to the `leading-log' approximation while   
$\mnt{i}^{(1)}$ is the final result including the `finite' corrections.
We note a difference of about 6~GeV in $\mu(M_{\rm SUSY})$ with respect  
to \spheno, which is however not of immediate concern for this 
analysis (it mainly stems from the different loop order of the 
effective Higgs potential  in the two programs). 
More important for us is the 2~GeV difference in the LSP pole mass 
due to the choice of scale for the self-energy corrections.

As a first step of improvement, we modify \isajet\ so that the 
one-loop self-energies for each neutralino (and also for each other sparticle) 
are computed with the renormalization scale set to its own mass scale.  
This means always setting the variable {\tt XLAM} in the routine 
{\tt SSM1LP} to the mass of the sparticle being renormalized. 
In this way, the double counting is much reduced. 
The result is shown as Case~B in \tab{mnt-isajet}. 
As expected, while the effect on the heavy neutralino masses is small,
there is a shift upwards of 2 GeV in the LSP mass. 

Next we invoke the complete step-beta functions 
of Ref. \cite{Dedes:1995sb} into the RGEs of {\it all} the SUSY parameters. 
The result of this is shown as Case~C in \tab{mnt-isajet}. 
As can be seen, the agreement with the other codes is less good 
in this case. One reason is that when integrating out one parameter, 
this leads in principle also to finite shifts for the parameters 
that remain in the RGEs. This is, however, not taken into account in 
the `naive' step-beta function approach, which employs continuous 
matching conditions. We will discuss this and other 
issues in more detail in \sect{discussion}.
 
As a consistency check, we also freeze out all SUSY parameters 
 at $Q=M_{\rm SUSY}$ in \isajetnn,  
and compute the pole masses at this scale 
analogous to what is done in the other codes. 
This is shown as Case D in \tab{mnt-isajet} and 
indeed gives the expected level of agreement with the other codes.

\section{Scalar top sector}

We next discuss the scalar top (stop) sector. 
At the tree level, the stop mass matrix is given by
\begin{equation}
  {\cal M}_{\st}^2 =
  \left( \begin{array}{cc}
     \mst{L}^2 & a_th_t \\ a_th_t & \mst{R}^2
  \end{array}\right) =
  \left( \begin{array}{cc}
  M_{\ti Q_3}^2+m_t^2+D_L & (A_t v_2-\mu v_1)h_t \\
  (A_t v_2-\mu v_1)h_t & M_{\ti U_3}^2+m_t^2+D_R
  \end{array}\right)\,,
  \label{eq:stmassmat}
\end{equation}
where $D_{L,R}$ denote the $D$-term contributions, 
$v_{1,2}$ are the Higgs VEVs and $h_t$ is the top Yukawa coupling. 
Using $h_t=m_t/v_2$, the off-diagonal element can also be written as 
\begin{equation}
  a_th_t= (A_t-\mu\cot\b)\,m_t \,,
  \label{eq:stoffdiag}
\end{equation}
with $m_t$ the running top-quark mass.
The difference to the neutralino sector is that this off-diagonal element 
can be large, actually as large or even larger than 
the diagonal elements.  
It can hence introduce a large mixing of $\st_L$ and $\st_R$, and  
much enhance the splitting of the mass eigenstates $\st_{1,2}$. 

This is not a problem in the common scale approach, where all 
the parameters are taken at the same scale and therefore 
\begin{equation}
  ({\cal M}_{\st}^2)^{\rm onshell} = 
  {\cal M}_{\st}^2(Q) + \D {\cal M}_{\st}^2(Q)\,, 
\label{eq:mstcorr}  
\end{equation}
with $\D {\cal M}_{\st}^2(Q)$ the one-loop self-energy corrections 
given in \cite{Pierce:1996zz}. 
Again, they have to be computed once for 
$\st_1$ and once for $\st_2$. This also results in a small 
${\cal O}(1\%)$ difference in the stop mixing angle, depending on whether  
it is taken at $p^2=\mst{1}^2$ or $p^2=\mst{2}^2$.

In the step-beta function approach applied in \isajet, 
the lowest-order 
mass matrix given in terms of $M_{\ti Q_3}^2(M_{\ti Q_3}^2)$,  
$M_{\ti U_3}^2(M_{\ti U_3}^2)$, $A_t(A_t)$, $\mu(M_{\rm SUSY})$, 
$v_{1,2}(M_{\rm SUSY})$ and $h_t(M_{\rm SUSY})$, and it becomes quite involved 
to define the corrections to the on-shell scheme in a consistent way. 
It is hence more convenient to integrate out all stop parameters 
at the average scale $Q_{\st}=M_{\rm SUSY}=\sqrt{\mst{L}\mst{R}}$ 
and add the self-energy 
corrections as in Eq.~\eq{mstcorr}. 

\Tab{mst-spheno} shows the results of \spheno\ for the stop sector at SPS1a 
for different renormalization scales $Q$. As can be seen, the scale dependence 
of the pole masses at one loop is about 1.5\%, that is ${\cal O}(\a_s^2)$.

\Tab{mst-isajet} shows the results 
of \isajetnn\ for Cases A--D, analogous to Sect.~3.3. 
Case B has a somewhat higher $\st_1$ mass than Cases A,D and \spheno. 
This can be explained by the fact that the $\st_1$ and $\st_2$ masses 
are not respectively evaluated at $p^2=\mst{1}^2$ and $p^2=\mst{2}^2$ 
in \isajet, but both at $p^2=\mst{L}^2$. 
Case C has somewhat lower top squark masses. The reason apparently is that
the gluino has been integrated out at $Q\simeq 600$ GeV, and its effect on
the squark soft masses is to {\it increase} 
their values during running from $M_{\rm GUT}$ to $m_{\st_{L,R}}$.

\begin{table}\begin{center}
\begin{tabular}{c||ccc|ccc|ccc}
  Q & $M_{\ti Q_3}$ & $M_{\ti U_3}$ & $A_t$ 
   & $\mst{1}^{^{\DR}}$ & $\mst{2}^{^{\DR}}$ & $\t_{\st}^{^{\DR}}$ 
   & $\mst{1}^{\rm pole}$ & $\mst{2}^{\rm pole}$ & $\t_{\st}^{\rm pole}$ \\
\hline
  100 & 533.8 & 454.6 & $-524.4$ & 414.1  & 607.0 & $56.5^\circ$
              & 395.8 & 577.0 & $56.4^\circ$ \\
  200 & 516.3 & 438.7 & $-508.9$ & 398.7 & 588.5 & $56.4^\circ$
              & 399.6 & 582.2 & $56.4^\circ$ \\
  $M_{\rm SUSY}$ & 495.5 & 419.9 & $-490.4$ & 380.2 & 566.2 &  $56.3^\circ$
              & 400.6 & 586.0 & $56.4^\circ$  \\
  1000 & 479.6 & 405.5 & $-476.3$ & 366.2 & 549.6 & $56.2^\circ$
              & 399.0 & 585.2 & $56.3^\circ$ \\
\hline
\end{tabular}\end{center}
\caption{\spheno\ results (masses in GeV) for the scalar top sector at SPS1a.  
}
\label{tab:mst-spheno}
\end{table}

\begin{table}\begin{center}
\begin{tabular}{c||ccc|ccc|ccc}
  Case & $M_{\ti Q_3}$ & $M_{\ti U_3}$ & $A_t$ 
   & $\mst{1}^{(0)}$ & $\mst{2}^{(0)}$ & $\t_{\st}^{(0)}$ 
   & $\mst{1}^{(1)}$ & $\mst{2}^{(1)}$ & $\t_{\st}^{(1)}$ \\
\hline
  A & 493.7& 422.5& $-496.6$ & 381.2& 566.4& $55.7^\circ$ & 401.8& 583.5& $56.8^\circ$ \\
  B & 493.6& 422.4& $-496.4$ & 381.2& 566.4& $55.7^\circ$ & 405.7& 584.9& $57.0^\circ$ \\
  C & 488.5& 414.3& $-489.7$ & 373.7 & 560.1 & $56.0^\circ$ & 394.1 & 577.2 & $57.2^\circ$ \\
\hline
  D & 495.3 & 420.6& $-498.5$ & 379.8& 566.6& $56.2^\circ$ & 400.0& 583.9& $57.2^\circ$ \\
\hline
\end{tabular}\end{center}
\caption{\isajetnn\ results (masses in GeV) for the scalar top sector at SPS1a.  
  Case A is the result of \isajet;
  Case B is \isajet\ with the improvement that the one-loop self-energies  
  are each computed at their relevant scale; 
  Case C employs step-beta functions for all SUSY parameters.   
  In case D, the SUSY parameters are all frozen out at 
  $Q=\sqrt{\mst{L}\mst{R}}\simeq 456$~GeV, and the one-loop corrections are 
  applied at this scale.}
\label{tab:mst-isajet}
\end{table}

\section{Improved Isajet: effect on the complete spectrum}

In order to improve on the problems discussed above, 
the new version of \isajetnn, {\tt v7.73}, adopts the following scheme: 
\begin{enumerate}
\item The running parameters of non-mixing sparticles, 
   i.e.\ squarks and sleptons 
   of the first and second generation, and the gluino, are extracted at 
   their respective mass scale. The one-loop radiative corrections are
   implemented at this scale, which is also taken to be the 
   renormalization scale. Thus, double counting of logarithmic
   corrections is not present.
\item The parameters of mixing sparticles, i.e.\ neutralinos, charginos, 
   stops, sbottoms, and staus, are all extracted at the scale 
   $Q=\sqrt{\mst{L}\mst{R}}$, which is also taken to 
   be the renormalization scale. Diagonalization of the mass matrices are
   performed once for each sparticle mass, with self-energies evaluated
   at the corresponding sparticle's tree-level mass.
\item The implementation of variable beta functions for the SUSY parameters 
   is postponed until a consistent treatment of logarithmic and finite 
   corrections for multiple scales is available.
\end{enumerate}
In addition, in \isajetNew, the gluino mass radiative corrections
depending on squark mixing have been added; these mixing corrections
were absent in previous \isajetnn\ versions. 
The results of \isajetNew\ are given in \tab{comp_sps1a} and 
compared with those of \isajet\ and \spheno. 
Also shown is the 
scale dependence in \spheno. 
\begin{table}\begin{center}
\begin{tabular}{c|ccc|c}
Mass    & \isajet & \isajetNew & \spheno & $\delta^{\rm scale}$\\
\hline
$\nt_1$       & 95.19 & 97.39 & 97.11 & 0.3\\
$\nt_2$       & 180.5 & 180.4 & 180.7 & 1.1\\
$\nt_3$       & 356.7 & 358.7 & 364.9 & 0.6\\
$\nt_4$       & 377.2 & 379.0 & 382.2 & 0.3\\
$\ch_1$       & 180.4 & 180.3 & 180.3 & 1.1 \\
$\ch_2$       & 376.2 & 378.0 & 383.3 & 0.4 \\
$\ti e_L$     & 203.2 & 203.3 & 202.4 & 0.3 \\
$\ti e_R$     & 144.0 & 142.5 & 144.1 & 0.8 \\
$\ti\nu_e$    & 187.1 & 185.5 & 186.2 & 0.2 \\
$\ti\tau_1$   & 134.8 & 134.6 & 134.4 & 0.6 \\
$\ti\tau_2$   & 206.7 & 205.9 & 206.4 & 0.3 \\
$\ti\nu_\tau$ & 186.2 & 183.4 & 185.3 & 0.2 \\
$\ti u_L$     & 559.5 & 564.9 & 565.1 & 9.8 \\
$\ti u_R$     & 544.0 & 548.6 & 547.8 & 8.9 \\
$\ti d_L$     & 565.2 & 570.9 & 570.5 & 9.7 \\
$\ti d_R$     & 543.7 & 548.2 & 547.8 & 8.9 \\
$\st_1$       & 401.8 & 395.2 & 400.6 & 5.0 \\
$\st_2$       & 583.5 & 584.4 & 586.0 & 9.0 \\
$\sb_1$       & 516.5 & 518.6 & 514.9 & 7.9 \\
$\sb_2$       & 539.7 & 547.1 & 547.5 & 8.7 \\
$\ti g$       & 611.4 & 605.9 & 604.3 & 1.3 \\
\hline
\end{tabular}\end{center}
\caption{Results of \isajetnn\ and \sphenonn\ for SPS1a. 
 \isajetNew\ is the new version, which includes the improvements 
 explained in the text; $\delta^{\rm scale}$ is the scale dependence 
 for $Q=0.1$--1~TeV in \spheno. All values in GeV.}
\label{tab:comp_sps1a}
\end{table}

At this point it is important to note that the remaining differences in the 
spectra of \isajetNew\ and \spheno\ 
cannot be attributed exclusively to the different methods of implementing 
SUSY thresholds to the SUSY parameters. 
There are analogous differences in the implementation of supersymmetric 
thresholds to the gauge and Yukawa couplings. 
In both programs, the experimental central values of the SM gauge couplings 
and third-generation fermion masses are used to determine the low-energy boundary 
conditions for the gauge and Yukawa couplings in the $\DR$ scheme. These are 
then extrapolated up to the GUT scale, defined as the scale where 
$\alpha_{1}= 5 \alpha_{e} /3(1- s_{W}^{2})$ and $\alpha_{2}= \alpha_{e}/s_{W}^{2}$
unify. 
\sphenonn, as well as \softsusynn\ and \suspectnn,  
use a one-step implementation of the supersymmetric (log+finite) corrections 
to gauge and Yukawa couplings with the matching between SM and MSSM couplings 
done at the scale $M_Z$. 
In \isajetnn, on the other hand, logarithmic thresholds
are implemented in a one-by-one decoupling from the RG equations 
each time a threshold is passed, while finite corrections are 
implemented collectively at a common scale. 
The boundary conditions at the GUT scale are therefore not going to be identical.
This is exemplified in Table \ref{tab:couplings}, where we show the GUT-scale 
values of the gauge and Yukawa couplings from \isajetnn\ and \sphenonn\ 
at SPS1a.
We see that there is no perfect 
gauge coupling unification since $\alpha_{3} \neq \alpha_{1} = \alpha_{2}$.
The unification degree is of the order of $1.5\%$. 
This is expected for two-loop RGEs and can be attributed to threshold 
effects due to particles with GUT scale masses. 
The differences between \isajetNew\ and \spheno\ amount to 
8\% for the GUT scale (which enters logarithmically in the RGEs),
$1\%$ for the unified gauge coupling, 
$2\%$ for the top and $8\%$ for the bottom Yukawa couplings.
We conclude that even for a `well behaved' point such as SPS1a
the differences in the GUT scale output cannot be neglected;  
in studies of Yukawa unified models they can become crucial.

\begin{table}\begin{center}
\begin{tabular}{c|ccc}
GUT output   & \isajet & \isajetNew & \spheno \\
\hline
$M_{\rm GUT}$       & $2.28\times 10^{16}$~{\rm GeV} & $2.28\times 10^{16}$~{\rm GeV}
   & $2.46\times 10^{16}$~{\rm GeV}   \\
$g_1(M_{G})=g_2(M_G)$ & 0.715 & 0.715 & 0.721 \\
$g_3 (M_{G}) $        & 0.706 & 0.706 & 0.707 \\
$h_{t}(M_{G})$        & 0.505 & 0.516 & 0.527 \\
$h_{b}(M_{G})$        & 0.049 & 0.047 & 0.051  \\
$h_{\tau}(M_{G})$     & 0.068 & 0.068 & 0.068 \\
\hline
\end{tabular}\end{center}
\caption{Comparison of $\DR$ GUT scale gauge and Yukawa couplings 
obtained from \isajet, \isajetNew\ and  \spheno\
for the benchmark point SPS1a.}
\label{tab:couplings}
\end{table}

\section{Discussion of the multiscale approach 
\label{sect:discussion}}

Integrating out all SUSY particles at a common scale is a reliable 
procedure if their masses are all in roughly the same ballpark. 
This is for example the case for the SPS1a benchmark point.   
However, in the case where the SUSY spectrum [including the Higgs bosons] 
is spread over a large range of masses, one should get more 
precise predictions 
by integrating out the SUSY particles at various scales. 
Consider for instance the gluino mass. In the common scale approach
\begin{equation}
   \msg^{\rm pole,CS} = M_3(Q)+\Delta\msg(Q)\,,
\end{equation}
while if the gluino is frozen out at its own mass scale, 
\begin{equation}
   \msg^{\rm pole,MS} = M_3(M_3)+\Delta\msg(M_3)\,.
   \label{eq:mgluinoMS}
\end{equation}
At SPS1a with $ M_{\ti Q}\sim 550$~GeV and $M_3\sim 600$~GeV, we find 
$\msg^{\rm pole,CS} = 604.3$~GeV and $\msg^{\rm pole,MS} = 604.1$~GeV, 
i.e.\ excellent agreement between the two methods.
If we move, however, $M_3$ to $M_3(M_3)=1.8$~TeV, we find 
$\msg^{\rm pole,CS} = 1512$~GeV,  
and $\msg^{\rm pole,MS} = 1531$~GeV 
--- i.e.\ a spread of 19 GeV. 
A multiscale treatment of thresholds therefore seems desirable 
when the spectrum is considerably split.

In the following, we outline the implications of a consistent 
multiscale approach. To work out a complete prescription 
is, however, beyond the scope of this Letter. 
Technically, at each scale a new effective theory (EFT) has to 
be constructed, and one faces the following difficulties:

\subsubsection*{1. Finite shifts} 

The naive way to take out particles from the RGEs via step functions
(as done e.g.~in \cite{Dedes:1995sb})
and to use continuous matching conditions for the remaining parameters
holds only for the lowest-order RGEs. At higher orders, finite shifts
have to be introduced, as has been known for a long time 
\cite{Weinberg:1980wa,Ovrut:1980dg,Ovrut:1980uv,Hall:1980kf,Binetruy:1980xn}. 
The simplest example is gauge coupling unification in $SU(5)$ theories: 
at lowest order, i.e.~using one-loop RGEs, the boundary conditions at 
$M_{\rm GUT}$ are given by 
$g_{\rm U(1)} = g_{\rm SU(2)} = g_{\rm SU(3)} = g_{\rm SU(5)}$. 
At next-to-leading and higher orders, finite threshold corrections due 
to particles with masses of order $M_{\rm GUT}$ have to be taken into account 
\cite{Weinberg:1980wa,Hall:1980kf,Binetruy:1980xn}, 
spoiling the lowest-order equality of the gauge couplings.
Another prominent example is the evolution of the strong coupling between
the scale of the lightest quarks and $m_Z$: see e.g.~\cite{Chetyrkin:1997un}.

In our case, we are working with two-loop (or higher loop) RGEs 
for the SUSY parameters, as for the gauge and Yukawa couplings. 
In this case, when integrating out SUSY particles at various scales, 
two issues have to be taken into account: \\
(i) The shifts of the $\DR$ parameters of the sparticle  
  that is integrated out to its pole-mass parameters involve
   contributions of all particles, which are degrees of freedom of the 
  current EFT. \\
(ii) The field(s) that are integrated out also lead to finite
  shifts in the boundary conditions of the parameters which remain in
  the RGEs. Take as an example the case where the gluino is the heaviest
  MSSM particle. At $Q=|M_3|$ the gluino can be integrated out and the
  pole gluino mass $m_{\tilde g}$ is obtained by taking into account
  the shifts from all strongly interacting particles (at the one-loop level; 
  at the two-loop level all particles contribute in principle). In addition,
  there will be finite shifts for $\alpha_s$, the Yukawa couplings (or 
  equivalently the masses) of the quarks, and the squark parameters.
  A similar effect has been known in QCD for a long time,  
  where the decoupling of heavy quarks
  also leads to shifts for the boundary conditions of running masses of
  the lighter quarks \cite{Wetzel:1981qg}.

\subsubsection*{2. New couplings} 

By integrating out part of the spectrum it may happen that the symmetry
of the EFT is `smaller' than the symmetry of the underlying theory.
In this case, there arise additional parameters in the EFT.
Moreover, there are in general higher-dimensional operators compatible
with the reduced symmetry; well known examples of this are the neutrino 
mass operator in the see-saw mechanism \cite{seesaw} 
or the operators governing rare meson/baryon decays.

In the case of supersymmetric theories, there is a well studied and 
important example, namely the Higgs sector. 
Within the MSSM, supersymmetry requires that three of the seven couplings
of a general 2-Higgs doublet model be zero and that the remaining four 
be expressed by gauge couplings (at least at tree level). Integrating
out the SUSY particles does not only lead to finite shifts for the four 
non-zero couplings, but also introduces non-zero values for the couplings
which are zero due to supersymmetry \cite{Haber:1993an}. In the very same
sector, a further complication arises if the top quark is integrated out, 
as $\rm SU(2)$ is broken in such a case. This leads to various  
new couplings to the $W$ and $Z$ bosons, each governed by its own
RGE \cite{Haber:1993an}.

A second important example in this context is the bottom Yukawa coupling.
If one integrates out, e.g.\ the gluino, from the spectrum, one induces 
non-holomorphic Yukawa couplings between the quark fields and the Higgs
fields, e.g.~a $b\bar{b}H_u$ coupling \cite{Carena:1999py}.

The complete set of superfield operators leading to dimension 5 and 6
operators in the Lagrangian has been worked out for the MSSM \cite{Piriz:1997id}, 
assuming that it is the effective theory of a more fundamental one.
If one regards the MSSM as fundamental and  integrates
out part of it, the operators obtained will form a subset of those 
given in \cite{Piriz:1997id}. The techniques to obtain the
tree-level coefficients in front of these operators have been elaborated
in \cite{Katehou:1987we}.

\subsubsection*{3. Sparticle mixing} 

Clearly, once from a set of mixing particles the heavier ones are
integrated out, the question arises of how to obtain  the mixing
effects that are visible in the case of a one-step decoupling. The
answer is that the above mentioned higher-dimensional operators 
introduce 
these effects.  
Let us sketch this for the stops, 
for the case when $\mst{L}^2 \gg \mst{R}^2$.
Integrating out  $\tilde t_L$ yields, for example, effective operators of the form
\begin{eqnarray}
&&c_1
 \frac{g A_t Y_t}{m^2_{\tilde t_L}} \tilde{t} P_R \tilde W_3  H^0_u \tilde t_R
\hspace{2mm} ,  \hspace{2mm}
c_2 \frac{- g \mu Y_t}{m^2_{\tilde t_L}}
 \tilde{t} P_R \tilde W_3  H^0_d \tilde t_R
\label{eq:effopt_neut}
\\
&&c_3
 \frac{|A_t Y_t|^2}{m^2_{\tilde t_L}} H^0_u {H^0_u}^*  \tilde t_R \tilde t^*_R
\hspace{2mm} ,  \hspace{2mm}
c_4 \frac{|\mu Y_t|^2}{m^2_{\tilde t_L}} H^0_d {H^0_d}^*  \tilde t_R \tilde t^*_R
\hspace{2mm} ,  \hspace{2mm}
c_5 \frac{- A_t \mu Y_t^2}{m^2_{\tilde t_L}}
 H^0_d {H^0_u}^*  \tilde t_R \tilde t^*_R + {\rm h.c.} .
\label{eq:effopt_mass}
\end{eqnarray}
After electroweak symmetry breaking, the operators of the  first line
contribute to the $\tilde t_1 t\tilde \chi^0_i$ vertices, whereas
those of the second line give contributions to the mass of the lighter stop.
The coefficients $c_i$ contain the information of the evolution of the
operators from the scale
$m_{\tilde t_L}$ down to $m_{\tilde t_1}$. As a check that these operators
really mimic the effect of mixing, let us calculate the stop mixing angle
for this case, taking into account that the left stop mass is much larger
than the $\mst{1}$ and $m_t$ (and thus the off-diagonal element):
\begin{equation}
\cos \theta_{\tilde t} =
 \frac{- m_t (A_t - \mu \cot \beta)}
      {\sqrt{ (m^2_{\tilde t_L} - m^2_{\tilde t_1})^2
            + m^2_t(A_t - \mu \cot \beta)^2 }} \simeq 
 \frac{- m_t (A_t - \mu \cot \beta)}{m^2_{\tilde t_L}} .
\end{equation}
At the scale $m_{\tilde t_L}$, $c_i=1\: \forall\: i$, and thus if one
naively replaces the Higgs fields by $v_{u,d} / \sqrt{2}$ and 
$\tilde t_R$ by $\tilde t_1$ one obtains
exactly the coupling of the lighter stop to the neutral wino in
Eq.~(\ref{eq:effopt_neut}) and the contribution of the left stop to
the lighter stop mass in Eq.~(\ref{eq:effopt_mass}). 
We note that the case of $c_i=1$ is also obtained if one
applies the see-saw formula to the stop sector.\\

\section{Conclusions}

We have discussed the two general methods of treating threshold effects in
the computation of sparticle pole masses from high scale input:
(a)~renormalization at a common SUSY scale and (b)~freezing out each sparticle
at its own mass scale. We have focused in particular on the concrete
implementations of these methods in \spheno\ and \isajet\
and compared the results of these programs for the SPS1a benchmark point.
Several shortcomings were pointed out, together with suggestions for improvements.
The new version \isajetNew\ incorporates these improvements, 
leading to better agreement between the various codes,
especially for the LSP mass but also 
for the squark and gluino masses.

Integrating out all SUSY particles at a common scale, as done in \softsusynn, 
\sphenonn\ and \suspectnn, is a reliable procedure if their masses are all 
in roughly the same ballpark. 
The results of the alternative method followed in \isajetnn\ are 
in good agreement with this procedure. 
(Here, note that the method applied in \isajetnn\ is not a complete 
multiscale but a hybrid approach.)
In order to match the ultra-high accuracy expected at the ILC, 
a consistent multiscale treatment of SUSY threshold effects  
seems necessary. 
The method is in principle well known from QCD. 
At each scale where a threshold is passed, the relevant field(s) have  
to be taken out of the RGEs and a new effective theory has to 
be constructed below the threshold. 
The shifts of the $\DR$ to the pole-mass parameters of the sparticle  
that is integrated out involve
contributions of all the particles that are degrees of freedom of the 
current EFT. Moreover, the field(s) that are integrated out 
lead to finite shifts in the boundary conditions of the parameters 
which remain in the EFT, so we face non-trivial matching conditions. 
Additional complications arise when the symmetry
of the EFT is `smaller' than the symmetry of the underlying theory.
We have outlined this multiscale approach in Sect.~6 and discussed 
its technical implications. 
A complete prescription for the matching over multiple scales 
is, however, beyond the scope of this Letter and 
will be explored in the future.

\section*{Acknowledgements}

The work of S.K. is financed by an APART (Austrian Programme of
Advanced Research and Technology) grant of the Austrian Academy of
Sciences. W.P.~is supported by a MCyT Ramon y Cajal contract, by the
Spanish grant BFM2002-00345, by the European Commission, Human
Potential Program RTN network HPRN-CT-2000-00148, and partly by the
Swiss `Nationalfonds'.


\end{document}